\documentclass[fleqn,10pt]{wlscirep}
\def\track#1{{\color{black}#1}}

\title{Local stability of cooperation in a continuous model of indirect reciprocity}

\author[1]{Sanghun Lee}
\author[2,$\ast$]{Yohsuke Murase}
\author[1,$\dagger$]{Seung Ki Baek}
\affil[1]{Department of Physics, Pukyong National University, Busan 48513, Korea}
\affil[2]{RIKEN Center for Computational Science, Kobe, Hyogo 650-0047, Japan}

\affil[$\ast$]{yohsuke.murase@gmail.com}
\affil[$\dagger$]{seungki@pknu.ac.kr}

\begin{abstract}
Reputation is a powerful mechanism to enforce cooperation among unrelated
individuals through indirect reciprocity, but it suffers from disagreement
originating from private assessment, noise, and incomplete information.
In this work, we investigate stability of cooperation in the donation
game by regarding each player's reputation and behaviour as continuous variables.
Through perturbative calculation, we derive a condition that a social norm
should satisfy to give penalties to its close variants, provided that everyone
initially cooperates with a good reputation, and this result is supported by
numerical simulation. A crucial factor of the condition is whether a
well-reputed player's donation to an ill-reputed co-player is appreciated by
other members of the society, and the condition can be reduced to a threshold
for the benefit-cost ratio of cooperation which depends on the reputational
sensitivity to a donor's behaviour as well as on the behavioural sensitivity to
a recipient's reputation.
Our continuum formulation suggests how indirect reciprocity can work
beyond the dichotomy between good and bad even in the
presence of inhomogeneity, noise, and incomplete information.
\end{abstract}
\begin{document}

\flushbottom
\maketitle

\thispagestyle{empty}

\section*{Introduction}

Reputation was an absolutely essential asset in trade of the illiterate in the
premodern era~\cite{burke1985day}, and it still plays a crucial role in markets
and communities, making reputation management a
central part of marketing and public relations.
Also in a variety of social contexts starting from early childhood, we
evaluate others based on third-party interactions~\cite{hamlin2007social}
and adjust our own behaviour to earn good reputations from
others~\cite{engelmann2012five}. In this regard, although some studies suggest
the existence of social evaluation in species other than
humans~\cite{abdai2016origin}, {\it Homo sapiens} seems to have unique
capability to use information of other social members through rumour and gossip.

Evolutionary biologists argue that the ability of social evaluation helps us
extend the range of cooperation beyond kinship
by encouraging cooperators and punishing defectors in a social dilemma~\cite{alexander1987biology,nowak1998evolution,leimar2001evolution,brandt2005indirect,nowak2005evolution,ohtsuki2009indirect,nax2015stability}.
A classical example of a social dilemma is the donation game, in which
a player's cooperation benefits his or her co-player by an amount of $b$ at
the cost of $c$, where $0<c<b$. The following payoff matrix defines the game:
\begin{equation}
\left(
\begin{array}{c|cc}
   & C & D\\\hline
C  & b-c & -c\\
D  & b & 0
\end{array}
\right),
\label{eq:payoff}
\end{equation}
where we abbreviate cooperation and defection as $C$ and $D$, respectively.
As is clearly seen from this payoff matrix, choosing $D$ is the rational choice
for each player whereas mutual cooperation is better for both, hence a
dilemma.
The players can escape from mutual defection by the action of
reciprocity if the game is repeated~\cite{axelrod1984evolution,baek2008intelligent,baek2016comparing,yi2017combination,murase2018seven,murase2020automata,murase2020five,murase2021friendly},
but the price is that they have to remember the past and repeat interaction with
sufficiently high probability, which is sometimes unfeasible.
The basic idea of indirect
reciprocity is that even a single encounter between two persons can be enough if
that experience is reliably transferred in the form of reputation to those who
will interact with these players in future.
\track{In other words, the problem is how to
store, transmit, and retrieve information on each others's past
behaviour in a distributed manner~\cite{nowak2005evolution,clark2020indirect}.
}
Experiments show that the notion of indirect reciprocity
provides a useful explanation for cooperative
human behaviour~\cite{wedekind2000cooperation,milinski2002reputation}.

For this mechanism to work, we need two rules as a social norm:
One is an assessment rule to assign reputation to a player based on his or her
action to another player. The other is a behavioural rule
to prescribe an action between $C$ and $D$, when players' reputations are
given.
An early idea was a norm called Image
Scoring, which judges the donor's $C$ and $D$ as good and bad,
respectively~\cite{nowak1998evolution}. According to this norm,
cooperation can thrive when
\begin{equation}
q > c/b,
\label{eq:image}
\end{equation}
where $q$ means the probability of knowing someone's
reputation~\cite{nowak2006five}. On the one hand, this condition seems natural
because it parallels Hamilton's rule for kin selection, and the only difference
is that $q$ has replaced genetic relatedness. On the other hand, if one asks
what is an essential prerequisite for a norm to promote cooperation, it is not
answered by Eq.~\eqref{eq:image}, and we need a broader perspective on the
structure of social norms.

According to Kandori's formalism~\cite{kandori1992social},
Image Scoring is an example of `first-order' assessment rules because its
judgment depends only on the donor's action. A `second-order' assessment rule
takes the recipient's reputation into account, and a `third-order'
assessment rule additionally refers to the donor's reputation.
The number of possible third-order rules thus amounts to $2^{2^3} = 256$. On the
other hand, the number of actions rules is $2^{2^2} = 16$ because a behavioural
rule prescribes an action depending on the reputations of the donor and
recipient. Among the $2^{2^3 + 2^2} = 4096$ combinations, we have
the {\em leading eight}~\cite{ohtsuki2004should,ohtsuki2006leading},
the eight pairs of an assessment rule $\alpha$ and a behavioural
rule $\beta$ that make cooperative equilibrium evolutionarily stable against
every mutant with $\beta' \neq \beta$ (Table~\ref{table:eight}).

\begin{table}
\caption{Leading eight. Cooperation and defection are denoted as $C$ and $D$,
respectively, and a player's reputation is either good ($1$) or bad ($0$).
By $\alpha_{uXv}$, we mean the reputation assigned to a player who
did $X \in \{C, D\}$ with reputation $u$ to another player with reputation $v$.
The behavioural rule $\beta_{uv}$ prescribes what a player should do between $C$
and $D$ when he or she has reputation $u$ and the co-player has reputation $v$.
We note that L1 has been known as Contrite Tit-for-Tat in the context of direct
reciprocity~\cite{sugden1986economics,boyd1989mistakes,panchanathan2003tale,takeuchi2007mathematics}.
}
\begin{tabular*}{\textwidth}{@{\extracolsep{\fill}}c@{~~}cccccccc@{~~~}cccc@{}}\hline
Rule & $\alpha_{1C1}$ & $\alpha_{1D1}$ & $\alpha_{1C0}$ & $\alpha_{1D0}$ &
 $\alpha_{0C1}$ & $\alpha_{0D1}$ & $\alpha_{0C0}$ & $\alpha_{0D0}$ &
 $\beta_{11}$ & $\beta_{10}$ & $\beta_{01}$ & $\beta_{00}$\\\hline
L1   & 1 & 0 & 1 & 1 & 1 & 0 & 1 & 0 & $C$ & $D$ & $C$ & $C$\\
L2 (Consistent Standing) & 1 & 0 & 0 & 1 & 1 & 0 & 1 & 0 & $C$ & $D$ & $C$ &
$C$\\
L3 (Simple Standing) & 1 & 0 & 1 & 1 & 1 & 0 & 1 & 1 & $C$ & $D$ & $C$ & $D$\\
L4  & 1 & 0 & 1 & 1 & 1 & 0 & 0 & 1 & $C$ & $D$ & $C$ & $D$\\
L5  & 1 & 0 & 0 & 1 & 1 & 0 & 1 & 1 & $C$ & $D$ & $C$ & $D$\\
L6 (Stern Judging) & 1 & 0 & 0 & 1 & 1 & 0 & 0 & 1 & $C$ & $D$ & $C$ & $D$\\
L7 (Staying) & 1 & 0 & 1 & 1 & 1 & 0 & 0 & 0 & $C$ & $D$ & $C$ & $D$\\
L8 (Judging) & 1 & 0 & 0 & 1 & 1 & 0 & 0 & 0 & $C$ & $D$ & $C$ & $D$\\\hline
\end{tabular*}
\label{table:eight}
\end{table}

The situation becomes complicated when reputations are not globally shared
in the population: Misjudgement does occur in the
presence of error, and some players may even have their own private rules
of
assessment~\cite{uchida2010effect,uchida2013effect,okada2017tolerant,okada2018solution}. Then, strict social norms such as `Judging' and `Stern Judging'
completely fail to tell if other players are good or bad, although they
successfully induce cooperation when reputation is always public
information~\cite{santos2018social,hilbe2018indirect}. Communication rounds can
be introduced to resolve disagreements~\cite{ohtsuki2009indirect}, or one may
need \track{empathy or} prudence in judgment to alleviate the
problem~\cite{radzvilavicius2019evolution,quan2020withhold}, but
these remedies imply the intrinsic instability of the reputation mechanism in
its pure sense. We also point out that most of the
existing models are based on an assumption that the dynamic variables
are binary, although reputation is not really a
simple dichotomy between good and bad,
and some actions cannot be classified as either cooperation or
defection~\cite{tanabe2013indirect,olejarz2015indirect}.

In this work, we thus wish to investigate indirect reciprocity by taking
reputations
and actions as continuous variables. By doing so, we can naturally deal with
the continuous dynamics between the existing norm and its close variants by
means of analytic tools.
We also expect that this formulation can be used to address the problems of
error and incompleteness: The idea is that perception error
will effectively \track{replace} a binary reputation by a probabilistic mixture between
good and bad, just as a binary action can be replaced by a probabilistic mixture
of cooperation and defection in the presence of implementation error. Although
the number of possible social norms expands to infinity, we will restrict
ourselves to local-stability analysis by assuming that mutants appear from a
small neighbourhood of the existing social norm.

\section*{Analysis}

Let us imagine a large population and denote the number of players
as $N$. The basic setting is that a random pair of players are picked up to
play the donation game [Eq.~\eqref{eq:payoff}].
In our model, the player chosen as a donor decides the degree of cooperation to
the co-player between zero and one, which mean full defection and full
cooperation, respectively, based on their reputations.
Let $m_{ij}$ denote player $j$'s reputation from the viewpoint of player $i$.
The player $i$ also has a behavioural rule $\beta_i (m_{ii}, m_{ij})$, which
determines how much he or she will do as a donor to $j$. Note that all of
$m_{ij}$, $\alpha_{i}$, and $\beta_i$ for any $i$ and $j$
take real values inside the unit interval.
Player $k$ is observing the interaction between $i$ and $j$, and it has
its own assessment rule $\alpha_k (m_{ki}, \beta_i, m_{kj})$. With observation
probability $q > 0$, the reputation that $k$ assigns to $i$ will be
updated on average as follows:
\begin{equation}
m_{ki}^{t+1} = (1-q) m_{ki}^t + \frac{q}{N-1} \sum_{j \neq i} \alpha_k \left[
m_{ki}^t, \beta_i \left(m_{ii}^t, m_{ij}^t \right), m_{kj}^t \right],
\label{eq:update}
\end{equation}
where the superscripts have been used as time indices.
\track{Equation~\eqref{eq:update} is to be analysed in this section.
Before proceeding, let us note two points:
First, as a deterministic equation, Eq.~\eqref{eq:update}
does not include error explicitly.
If the probability of error is low, Eq.~\eqref{eq:update}
will nevertheless describe the dynamics for most of the time, and the main
effect of error will be to perturb the output of $\alpha$ or $\beta$
by a small amount at a point in time, say, $t=0$.
Second, from a mathematical point of
view, it is preferable to treat both diagonal and off-diagonal elements on an
equal footing as in Eq.~\eqref{eq:update}, which implies that
one has to observe even the self-reputation $m_{ii}$ probabilistically.
If that sounds unrealistic,
we may alternatively assume that donors and recipients
update their self-reputations with probability one. However, it is a reasonable
guess that the difference between these two settings becomes marginal when $N$
is large enough, and this guess is indeed verified by numerical calculation (not
shown).
}

Throughout this work, $\alpha$ and $\beta$ are assumed to be
C$^2$-differentiable. \track{In addition, we will focus on the cases where
the system has a fixed point characterized by
%\begin{linenomath}
\begin{subequations}
\begin{align}
\alpha(1,1,1) &=1\\
\beta(1,1)&=1
\end{align}
\label{eq:fp}
\end{subequations}
%\end{linenomath}
because otherwise the norm would not sustain cooperation
among well-reputed players from the start.}
As concrete examples of $\alpha$ and $\beta$,
let us extend the leading eight to deal with
continuous variables by applying the trilinear (bilinear) interpolation to
$\alpha$ ($\beta$) in Table~\ref{table:eight}.
If we consider L3 (Simple Standing), for instance, it is described by
%\begin{linenomath}
\begin{subequations}
\begin{align}
\alpha_\text{SS}(x,y,z) &= yz - z + 1\\
\beta_\text{SS}(x,y) &= y.
\end{align}
\label{eq:ss0}
\end{subequations}
%\end{linenomath}
If we define \track{$A_\xi \equiv \left. \partial \alpha / \partial \xi
\right|_{(1,1,1)}$ and $B_\lambda \equiv \left. \partial \beta / \partial
\lambda \right|_{(1,1)}$} with $\xi \in \{x,y,z\}$ and $\lambda \in \{x,y\}$,
all the leading eight have \track{$A_y = B_y = 1$}, together with \track{$A_x =
B_x = 0$}, and these are related with the basic properties of the leading
eight to be nice, retaliatory, apologetic, and
forgiving~\cite{ohtsuki2006leading}.

\track{
Below, we will examine two aspects of stability: The first is recovery of full
cooperation from disagreement in a homogeneous
population where everyone uses the same $\alpha$ and
$\beta$~\cite{hilbe2018indirect}.
Starting from $m_{ij}=1$ for every $i$ and $j$,
the dynamics of Eq.~\eqref{eq:update} will be investigated within the
framework of linear-stability analysis. The
second aspect is the stability against mutant norms, for which
we have to check the long-term payoff difference between the resident and mutant
norms in a stationary state.
We again start this
analysis from a nearly homogeneous population in which only one individual
considers using a slightly different norm.
Although private assignment of reputation is allowed, the point is that
it will remain unrealised if no one has a
reason to deviate from the prevailing norm, considering that such deviation
will only decrease his or her own payoff.
In this sense, the homogeneity serves as a self-consistent
assumption in the second part of the stability analysis.
}

\subsection*{Recovery from disagreement}

To understand the time evolution of disagreement in a homogeneous population
with common $\alpha$ and $\beta$, let us rewrite Eq.~\eqref{eq:update}:
\begin{equation}
m_{ki}^{t+1} = (1-q) m_{ki}^t + \frac{q}{N-1} \sum_{j \neq i} \alpha \left[
m_{ki}^t, \beta \left( m_{ii}^t, m_{ij}^t \right), m_{kj}^t \right],
\end{equation}
where $\alpha_k = \alpha$ and
and $\beta_i = \beta$ in this homogeneous population.
\track{Initially, everyone starts with a good reputation, which can be
perturbed by error. To see whether the magnitude of the perturbation grows with
time, we set $m_{ki}^t \equiv 1-\epsilon_{ki}^t$ and expand the above equation
to the first order of $\epsilon$ as follows:}
\track{
\begin{eqnarray}
1-\epsilon_{ki}^{t+1} &=& (1-q) \left(1-\epsilon_{ki}^t \right) + \frac{q}{N-1}
\sum_{j \neq i}
\alpha \left[ 1-\epsilon_{ki}^t, \beta\left( 1-\epsilon_{ii}^t,
1-\epsilon_{ij}^t \right), 1-\epsilon_{kj}^t \right]\\
&\approx& (1-q) \left(1-\epsilon_{ki}^t \right) + \frac{q}{N-1}
\sum_{j \neq i}
\alpha \left[ 1-\epsilon_{ki}^t, 1- \left(B_x\epsilon_{ii}^t + B_y
\epsilon_{ij}^t \right),
1-\epsilon_{kj}^t \right]\\
&\approx& (1-q) \left(1-\epsilon_{ki}^t \right) + \frac{q}{N-1}
\sum_{j \neq i}
\left\{
1- \left[ A_x \epsilon_{ki}^t + A_y \left( B_x\epsilon_{ii}^t + B_y
\epsilon_{ij}^t \right) + A_z \epsilon_{kj}^t \right] \right\},
\end{eqnarray}
or, equivalently,
}
\begin{eqnarray}
\epsilon_{ki}^{t+1} &\approx& (1-q) \epsilon_{ki}^t + \frac{q}{N-1}
\sum_{j \neq i} [\track{A_x} \epsilon_{ki}^t + \track{A_y} (\track{B_x} \epsilon_{ii}^t +
\track{B_y} \epsilon_{ij}^t) + \track{A_z} \epsilon_{kj}^t]\\
&=& (1-q + q\track{A_x}) \epsilon_{ki}^t + q \track{A_y} \track{B_x} \epsilon_{ii}^t +
\frac{q}{N-1}
\sum_{j \neq i} [\track{A_y} \track{B_y} \epsilon_{ij}^t + \track{A_z} \epsilon_{kj}^t],
\label{eq:recovery}
\end{eqnarray}
which leads to
\begin{eqnarray}
\frac{d}{dt} \epsilon_{ki} \approx
-q(1-\track{A_x}) \epsilon_{ki} + q \track{A_y} \track{B_x} \epsilon_{ii} +
\frac{q}{N-1}
\sum_{j \neq i} [\track{A_y} \track{B_y} \epsilon_{ij} + \track{A_z} \epsilon_{kj}],
\end{eqnarray}
if time is regarded as a continuous variable.
This is a linear-algebraic system with an $N^2 \times N^2$ matrix. In principle,
we can find the stability at the origin as well as the speed of convergence
toward it by calculating the eigenvalues.
\track{By attempting this calculation from $N=2$ to $5$ with a symbolic-algebra
system~\cite{Mathematica}, we see the following pattern in the eigenvalue
structure:}
\begin{eqnarray}
\Lambda_1^{(N^2-2N+1)} &=& q\left(-1+\track{A_x}- \frac{1}{N-1}\track{A_z} \right)\\
\Lambda_2^{(N-1)} &=& q(-1+\track{A_x}+\track{A_z})\\
\Lambda_3^{(N-1)} &=& q\left(-1+\track{A_x}-\frac{1}{N-1}\track{A_z}+\track{A_y} \track{B_x}
-\frac{1}{N-1}\track{A_y} \track{B_y} \right)\\
\Lambda_4^{(1)} &=& q(-1 + \track{A_x} + \track{A_z} + \track{A_y} \track{B_x} + \track{A_y}
\track{B_y}),
\end{eqnarray}
where each superscript on the left-hand side
means multiplicity of the corresponding eigenvalue.
\track{Based on this observation, we conjecture that this pattern is valid for
general $N$.}
A sufficient condition for recovery to take place in this first-order
calculation is that the largest eigenvalue is negative. The largest eigenvalue
is the last one, $\Lambda_4^{(1)}$, because all the derivatives are
non-negative. In other words, the first-order
perturbation analysis gives a sufficient condition for local recovery as
\begin{equation}
Q \equiv -1 + \track{A_x} + \track{A_z} + \track{A_y} (\track{B_x} + \track{B_y}) < 0.
\label{eq:Q}
\end{equation}

\subsection*{Suppression of mutants}
\track{To analyse the effect of a mutant norm, we will look at the long-time
behaviour in Eq.~\eqref{eq:update}. That is,}
for given sets of rules $\{ \alpha_i \}$ and $\{ \beta_i \}$, we assume that
the image matrix $\{ m_{ij} \}$ will converge to a stationary state as $t \to
\infty$, satisfying
\begin{equation}
m_{ki} = \frac{1}{N-1} \sum_{j \neq i} \alpha_k \left[
m_{ki}, \beta_i (m_{ii}, m_{ij}), m_{kj} \right].
\label{eq:stationary}
\end{equation}
Note that $q$ only affects the speed of convergence to stationarity:
It is an irrelevant parameter as far as we work with a stationary state,
which is in contrast with Eq.~\eqref{eq:image},
where $q$ appears as an essential condition for indirect reciprocity.
In the donation game with benefit $b$ and cost $c$ [Eq.~\eqref{eq:payoff}],
player $j$'s expected payoff can be computed as
\begin{equation}
\pi_j = \frac{1}{N-1} \left[ b \sum_{i \neq j} \beta_i (m_{ii}, m_{ij}) - c
\sum_{i \neq j} \beta_j (m_{jj}, m_{ji}) \right].
\end{equation}
For the sake of simplicity, let us assume that every person with index $1$ to
$N-1$ has the same rules and equal reputation, so that player $i=1$ is
representative for all of them in the resident population. Now, the situation is
effectively reduced to a two-body problem between players $0$ and $1$.
By assumption, the system initially starts from a fully cooperative state
where everyone has good reputation, i.e., $m_{11} = \beta(1,1) = \alpha(1,1,1)=
1$. The rules used by the resident population will be denoted by $\alpha \equiv
\alpha_1$ and $\beta \equiv \beta_1$ without the subscripts.
Now, the focal player $0$ attempts a slightly different norm,
defined by $\alpha_0 (x,y,z) = \alpha(x,y,z) - \delta(x,y,z)$ and
$\beta_0(x,y) = \beta(x,y) - \eta(x,y)$ with
$|\delta| \ll 1$ and $|\eta| \ll 1$.
\track{Let us assume that the introduction of $\delta$ and $\eta$ causes small
changes in the image matrix: Only the elements related to the focal player will
be affected because the residents can still give $m_{11}=1$ to each other
when the mutant occupies a negligible fraction of the population,
i.e., $N \gg 1$. Therefore,
if mutation leads to} $m_{00} = 1-\epsilon_{00}$, $m_{01} = 1-\epsilon_{01}$, and
$m_{10} = 1-\epsilon_{10}$ with $\epsilon_{ij} \ll 1$,
by expanding Eq.~\eqref{eq:stationary} to the linear order of
perturbation (see Methods), we obtain
\begin{eqnarray}
\epsilon_{00} &=& \frac{(1-\track{A_x}+\track{A_y} \track{B_y})\delta_1 +
(1-\track{A_x}-\track{A_z})\track{A_y} \eta_1}{(1-\track{A_x}-\track{A_z})(1-\track{A_x}-\track{A_y}
\track{B_x})}\label{eq:epsilon00}\\
\epsilon_{01} &=& \frac{\delta_1}{1-\track{A_x}-\track{A_z}}\label{eq:epsilon01}\\
\epsilon_{10} &=& \frac{(\track{B_x}+\track{B_y})\delta_1 +
(1-\track{A_x}-\track{A_z})\eta_1}{(1-\track{A_x}-\track{A_z})(1-\track{A_x}-\track{A_y} \track{B_x})}
\track{A_y},\label{eq:epsilon10}
\end{eqnarray}
where $\delta_1 \equiv \delta(1,1,1) \ge 0$ and $\eta_1 \equiv \eta(1,1) \ge 0$,
provided that
\begin{eqnarray}
\track{A_x} + \track{A_z} &<& 1 \label{eq:ineq1}\\
\track{A_x} + \track{A_y} \track{B_x} &<& 1. \label{eq:ineq2}
\end{eqnarray}
We can now calculate the focal player $0$'s payoff as follows:
\begin{eqnarray}
\pi_0 &=& \frac{1}{N-1} \left[ b \sum_{i \neq 0} \beta_i (m_{ii}, m_{i0}) - c
\sum_{i \neq 0} \beta_0 (m_{00}, m_{0i}) \right]\\
&=& b \beta(m_{11}, m_{10}) - c\beta_0 (m_{00}, m_{01})\\
&\approx& b \left( 1- \track{B_y} \epsilon_{10} \right) - c \left( 1-\track{B_x}
\epsilon_{00} - \track{B_y} \epsilon_{01} - \eta_1 \right).
\end{eqnarray}
If we plug Eqs.~\eqref{eq:epsilon00}, \eqref{eq:epsilon01}, and
\eqref{eq:epsilon10} here,
the payoff change $\Delta \pi_0 \equiv \pi_0 - (b-c)$ is given as
\begin{eqnarray}
\Delta \pi_0 = -\frac{b\track{A_y} \track{B_y} - c(1-\track{A_x})}{1-\track{A_x} -\track{A_y}
\track{B_x}} \left[ \left( \frac{\track{B_x}+\track{B_y}}{1-\track{A_x}-\track{A_z}} \right)
\delta_1 + \eta_1 \right],
\label{eq:dpi0}
\end{eqnarray}
and we require this quantity to be negative
for any small positive $\delta_1$ and $\eta_1$.
\track{
Here, it is worth stressing that
the signs of $\delta_1$ and $\eta_1$ are determined because
we start from a fully cooperative state with $m_{ij}=1$:
For other states where $\delta$ and
$\eta$ can take either sign, the first-order terms should
vanish so that the second-order terms can determine the sign of $\Delta \pi_0$.
In this respect, the payoff analysis is greatly simplified by choosing the
specific initial state.
}
Because of Eqs.~\eqref{eq:ineq1} and \eqref{eq:ineq2}, the
negativity of Eq.~\eqref{eq:dpi0} is reduced to the following inequality:
\begin{equation}
\frac{b}{c} > \frac{1-\track{A_x}}{\track{A_y} \track{B_y}},
\label{eq:bc}
\end{equation}
which, together with Eqs.~\eqref{eq:ineq1} and \eqref{eq:ineq2}, characterizes
a condition for a social norm to stabilize cooperation against local mutants,
as an alternative to Eq.~\eqref{eq:image}.
This result is
intuitively plausible because cooperation will be unstable if
one does not lose reputation by decreasing the degree of cooperation (i.e.,
$\track{A_y} \approx 0$) or if no punishment is imposed on an ill-reputed player
(i.e., $\track{B_y} \approx 0$).

Two remarks are in order: First, whether mutation occurs to a single individual
or to a fraction of the population does not alter the final
result \track{in this first-order calculation}. Suppose that the population is divided into two groups with fractions
$p$ and $1-p$, respectively. One group has $\alpha$ and $\beta$, and the other
group has $\alpha+\delta$ and $\beta+\eta$. Then, the payoff difference between
two players, each from a different group, is still the same as
Eq.~\eqref{eq:dpi0} (see Methods).
Therefore, if an advantageous mutation occurs with $p \ll 1$, the mutants are
always better off than the resident until they take over the whole population,
i.e., $p \to 1$. In this sense, our condition determines not only the initial
invasion but also the fixation of a mutant norm, as long as it is a
close variant of the resident one.
Second, one could ask what happens if a mutant differs only
in the slopes while keeping $\delta_1=\eta_1=0$. Equation~\eqref{eq:dpi0} does
not answer this question because it is based on an assumption that the
$\left. \partial \delta / \partial \xi \right|_{(1,1,1)} \epsilon_{ij}$ and
$\left. \partial \eta / \partial \lambda \right|_{(1,1)} \epsilon_{ij}$, where
$\xi \in \{ x, y, z\}$ and $\lambda \in \{x, y\}$, are all negligibly small in
the first-order calculation.
However, even if the derivatives are taken into
consideration, we find that $\delta_1$
or $\eta_1$ must still be positive to make a finite payoff change.
In other words, the basic form of Eq.~\eqref{eq:dpi0} is still useful, although
the coefficients include correction terms. The performance of such a `slope
mutant' will be checked numerically at the end of the next section.

\section*{Results}

\track{In this section, we will numerically check the continuous-reputation
system in the presence of inhomogeneity, noise, and incomplete information. More
specifically, the simulation code should allow each player $i$ to adopt a
different set of $\alpha_i$ and $\beta_i$ to simulate an inhomogeneous
population. The outputs of $\alpha_i$ and $\beta_i$ can be affected by
random-number generation to simulate a noisy environment where misperception and
misimplementation occur, and every interaction between a pair of
players will update only some part of the reputation system, parametrized by
the observation probability $q$, because information is incomplete.
}

Our numerical simulation code is based on a publicly available
one~\cite{hilbe2018indirect} but has been modified to handle continuous
variables. To simulate the dynamics of a society of $N$ players,
we work with an $N \times N$ image matrix $\{ m_{ij} \}$ whose elements are all
set to be ones at the beginning. Every player starts with zero payoff, i.e.,
$\pi_i = 0$ initially. In each round, we randomly pick up two players,
say, $i$ and $j$, so that $i$ is the donor and $j$ is the recipient of the
donation game [Eq.~\eqref{eq:payoff}], which has $b=2$ and $c=1$ unless
otherwise noted. Each other member of the population,
say, $k$, independently observes the interaction with probability $q$ and
updates $m_{ki}$ according to his or her own assessment rule $\alpha_k$.
Although the above analyses are generally applicable to any norms defined by
$\alpha$ and $\beta$ \track{as long as Eq.~\eqref{eq:fp} is true}, we would like
to focus on Simple Standing as a representative example of successful norms.
\track{Misperception may occur with probability $e$, whereby
$m_{ki}$ becomes a random number drawn from the unit interval.
Implementation error is also simulated in a similar way
by setting the output of $\beta$ to a random number between zero (defection) and
one (cooperation) with probability $\gamma$.}
This process is repeated for $M$ rounds, during which every player's payoff is
accumulated. Equation~\eqref{eq:stationary} suggests that $q$ will only affect
the convergence rate toward a stationary state. For this reason, we will fix
this parameter at $q=0.4$ throughout the simulation unless otherwise mentioned.
Note also that we have deliberately \track{made} this parameter low enough to violate
the inequality in Eq.~\eqref{eq:image}.

\begin{figure}
\begin{center}
\includegraphics[width=0.45\textwidth]{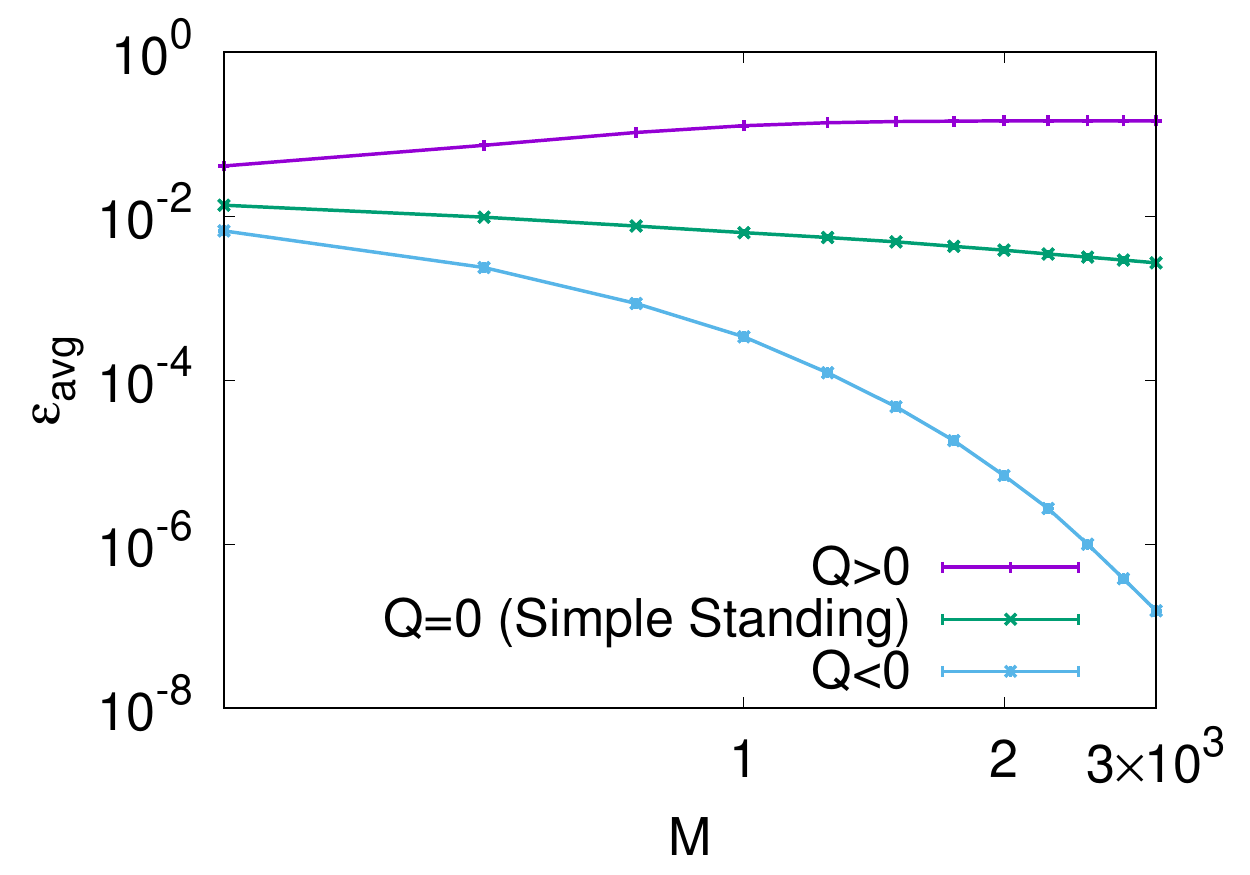}
\end{center}
\caption{Recovery from disagreement when $M$ rounds have elapsed
in a population of size $N=50$ with common $\alpha$ and $\beta$.
Initially, we randomly pick up $20\%$ of the image-matrix elements and change
them to $0.9$, whereas the rest of them remain as $1$'s, and the simulation has
been repeated over $10^3$ independent samples \track{without error, i.e.,
$e=\gamma=0$}.
In this log-log plot,
the vertical axis shows the average difference from the state of perfect
reputation, represented by the average of $\epsilon_{ki} \equiv 1-m_{ki}$.
We have tested three norms, which all have $\alpha(1,1,1)=1$ and $\beta(1,1)=1$
but differ in their local slopes there:
The first norm has $(\track{A_x}, \track{A_y}, \track{A_z}) = (0.2,0.9,0.1)$ and
$(\track{B_x}, \track{B_y}) = (0.2,0.8)$, which together yield $Q \equiv -1 + \track{A_x} +
\track{A_z} + \track{A_y} (\track{B_x} + \track{B_y})>0$ [Eq.~\eqref{eq:Q}].
The next one is Simple Standing with $(\track{A_x}, \track{A_y}, \track{A_z}) =
(1,0,1)$ and $(\track{B_x}, \track{B_y}) = (0,1)$, which has $Q=0$.
The last one for $Q<0$ is a variant of Simple Standing with $(\track{A_x},
\track{A_y}, \track{A_z}) = (0,0.9,0)$ and $(\track{B_x}, \track{B_y}) = (0,0.9)$.
}
\label{fig:recovery}
\end{figure}

To see the effect of $Q$ on recovery [Eq.~\eqref{eq:Q}], we have tested three
norms one by one in a homogeneous population \track{with $e=\gamma=0$} (Fig.~\ref{fig:recovery}).
All these norms have $\alpha(1,1,1)=1$ and $\beta(1,1)=1$ in common but their
local slopes are different to make $Q$ positive, zero, or negative.
The first norm under consideration
has $(\track{A_x}, \track{A_y}, \track{A_z}) = (0.2,0.9,0.1)$
and $(\track{B_x}, \track{B_y}) = (0.2,0.8)$, which together make $Q>0$.
If some members of
the population initially have slightly imperfect reputations, they fail to
recover under such a norm.
If $Q<0$, on the other hand, the recovery process indeed
takes place with a finite time scale. Although Simple Standing violates
Eq.~\eqref{eq:Q} by having $Q=0$, our simulation shows that it gets reputation
recovered with the aid of
higher-order terms, and it is a slow process with a diverging time scale.
Among the leading eight, L1, L3 (Simple Standing), L4, and L7 (Staying)
fall into this category of $Q=0$, whereas the other four, i.e., L2 (Consistent
Standing), L5, L6 (Stern Judging), and L8 (Judging), have positive $Q$.
The difference between these two groups is whether $\track{A_z} = \alpha_{1C1} -
\alpha_{1C0} = 1-\alpha_{1C0}$ is zero or one: If a well-reputed player has to
risk his or her own reputation in helping an ill-reputed co-player, i.e.,
$\alpha_{1C0}=0$, it means $\track{A_z}=1$ and $Q>0$, so we can conclude that
the initial state of $m_{ki} \approx 1$ will not be recovered.
According to an earlier study on the leading eight~\cite{hilbe2018indirect}, the
latter four with
$Q>0$ have long recovery time from a single disagreement in reputation.
Although it is not
derived from a continuum formulation, the result is qualitatively consistent
with ours.

\begin{figure}
\begin{center}
\includegraphics[width=0.45\textwidth]{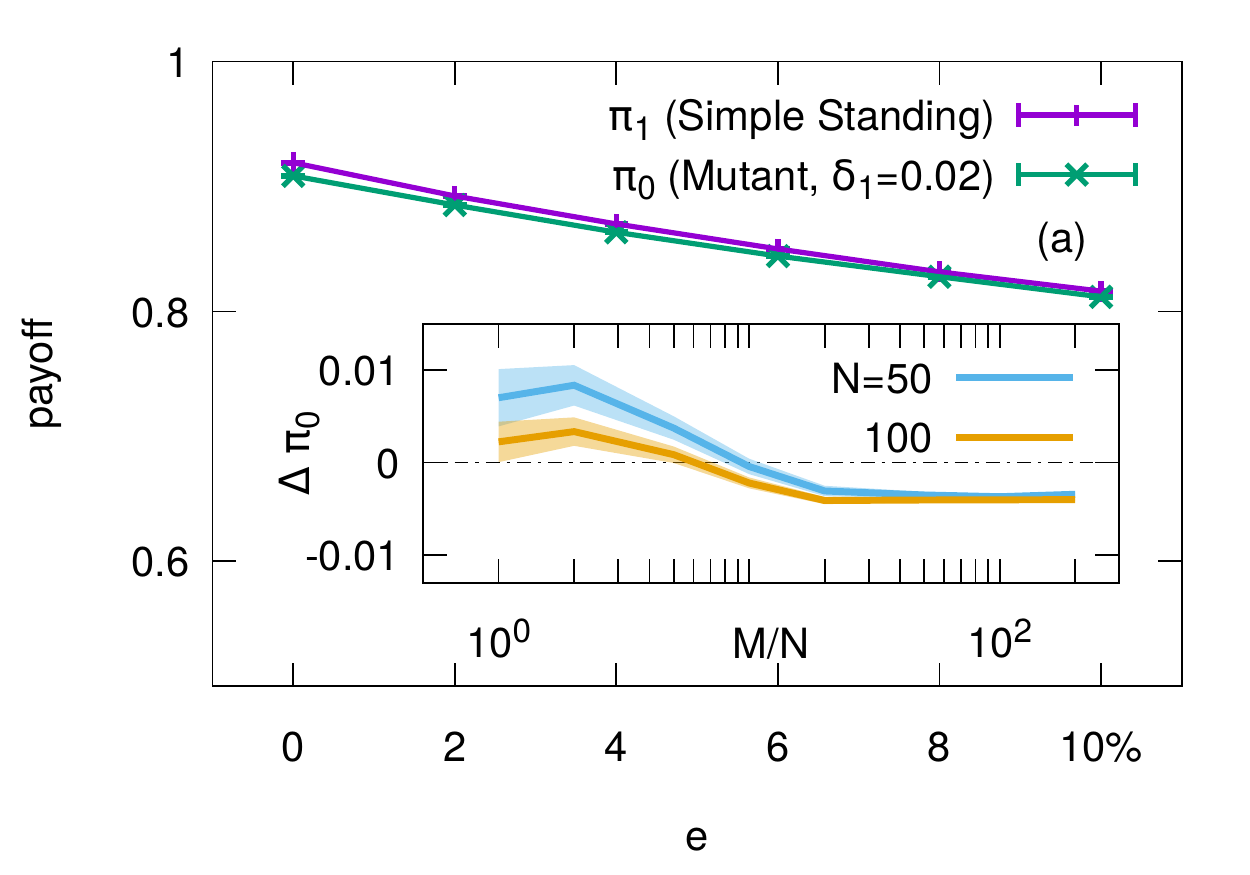}
\includegraphics[width=0.45\textwidth]{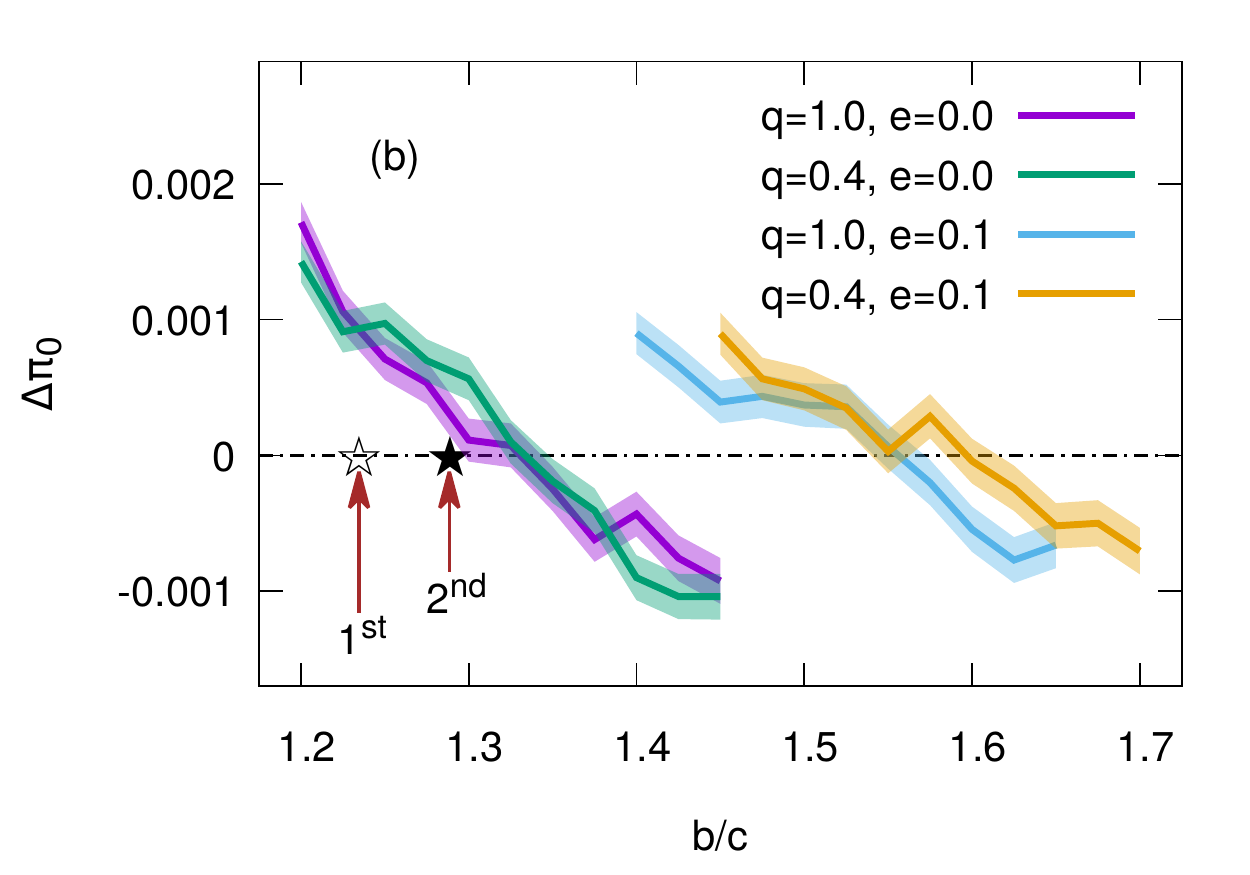}
\end{center}
\caption{
Stationary states of a population with $N=50$ players, reached from
an initial condition with $m_{ij}=1$ for every $i$ and $j$. In each case,
the mutant norm differs from the resident one by $\delta_1 = 0.02$ and occupies
one half of the population ($p=0.5$). \track{
The game is defined by Eq.~\eqref{eq:payoff} with $b=2$ and $c=1$.}
(a) Average payoffs over \track{$5\times$}$10^4$ samples when the resident norm is Simple
Standing.
Everyone can observe each interaction with probability $q=0.4$, and
\track{perception error and implementation error occur with probabilities $e =
0.1$ and $\gamma = 0.1$, respectively.}
Inset: Convergence of payoff difference $\Delta \pi_0 \equiv \pi_0 - \pi_1$
\track{as $M$ increases}.
If $M \propto gN$ with a sufficiently large constant $g \gtrsim O(10)$,
the mutants will obtain less payoffs
than Simple Standing, making $\Delta \pi_0 < 0$.
This result has no significant dependence on $N$.
(b) Payoff advantage of mutants with respect to the resident as a function
of $b/c$, averaged over \track{$5\times 10^4$} samples per each, \track{when
$M=10^4$}.
The resident norm, a variant of Simple Standing, has
$\alpha(1,1,1)=\beta(1,1)=1$ and $\track{A_x}
= \track{A_z} = \track{B_x} = 0$ but $\track{A_y} = \track{B_y} = 0.9$
as in Fig.~\ref{fig:recovery}. \track{Implementation error occurs with
probability $\gamma=0.1$, and the results are qualitatively the same for
any small $\gamma$.}
The stars on the horizontal line
indicate the predicted threshold values obtained from the first-order and
second-order calculations, respectively.
\track{In both of these panels, the shaded areas represent error bars.}
}
\label{fig:half}
\end{figure}

As for the effect of mutation in assessment rules,
let us consider the following scenario: One half of the population
have adopted Simple Standing [Eq.~\eqref{eq:ss0}],
whereas the other half are ``mutants'' using a different assessment rule
$\alpha_\text{SS} - \delta$ with
\begin{equation}
\delta(x,y,z) = \delta_1 (2yz - 2z + 1),
\label{eq:delta}
\end{equation}
where $\delta_1$ is a small number, say, $\delta_1 = 0.02$ in numerical
calculation.
Such a half-and-half configuration is being used because the payoff
difference [Eq.~\eqref{eq:dpi0}] is unaffected by the fraction of mutants, $p$
(see Methods).
Figure~\ref{fig:half}(a) shows that the level of cooperation is still high
if $e \ll 1$, and the cooperation rate of Simple Standing
in the continuous form
converges to $100\%$ in a monomorphic population (not shown).
Furthermore, we see that mutants are worse off than the players of
Simple Standing, i.e., $\pi_0 < \pi_1$, as expected.

From a theoretical viewpoint,
an important question is how quickly the mutants' payoff difference $\Delta
\pi_0 \equiv \pi_0 - \pi_1$
becomes negative: Although we have argued that the inequality will be true for
Simple Standing, the calculation is based on several assumptions. In particular,
one could say that Eq.~\eqref{eq:update} corresponds to $M \propto N^2$ because
it seems to assume that everyone
meets every other player with a weighting factor of $1/(N-1)$. If $M \propto
N^2$, however, it would pose a serious obstacle to applying such a norm to the
society where the number of interactions will grow linearly with $N$.
Fortunately, the inset of Fig.~\ref{fig:half}(a) shows that $M \propto N$ indeed
suffices to make $\Delta \pi_0$ negative.
One could also point out that the
payoff difference should be $\Delta \pi_0 = - \delta_1$ according to
Eq.~\eqref{eq:dpi0}, whereas the result in Fig.~\ref{fig:half}(a) has smaller
magnitude. A part of the reason is that Eq.~\eqref{eq:dpi0} does not take
perception error into account,
so the numerical value recovers the predicted order of
magnitude as $e \to 0$. In addition, Eq.~\eqref{eq:dpi0}
is based on a first-order approximation, and a higher-order calculation
reproduces the observed value with greater precision (see Methods).

\track{An important prediction of our analysis is the threshold of $b/c$ to make
a local mutant worse off than the resident population [Eq.~\eqref{eq:bc}].}
In Fig.~\ref{fig:half}(b), we directly check Eq.~\eqref{eq:bc} by measuring
payoffs in equilibrium in a population of size $N=50$.
A variant of Simple Standing is chosen as the resident norm, which occupies
$p=0.5$ of the population with $\alpha(1,1,1)=\beta(1,1)=1$ and $\track{A_x} =
\track{A_z} = \track{B_x} = 0$. The only difference from Simple Standing is that
$\track{A_y} = \track{B_y} = 0.9$, and the reason of this variation is that the
first-order perturbation for the leading eight develops spurious singularity
when $p$ is finite (see Methods). When perception is free from error, i.e.,
$e=0$, the results do not depend on the observation probability $q$,
as expected from stationarity [Eq.~\eqref{eq:stationary}],
and the threshold value is consistent with the first- and second-order
calculations [the arrows in Fig.~\ref{fig:half}(b)]. When $e>0$,
on the other hand, the threshold is pushed upward, implying that cooperation
becomes harder to stabilize because of the perception error. In addition, we now
see that incomplete information with $q<1$ can shift the threshold further
with the aid of positive $e$.
\track{We have also changed the value of $\gamma$, but it does not
not change the average behaviour in the above results.
Overall, the point of
Fig.~\ref{fig:half}(b) is that our analysis does capture the correct picture.}

\begin{figure}
\begin{center}
\includegraphics[width=0.45\textwidth]{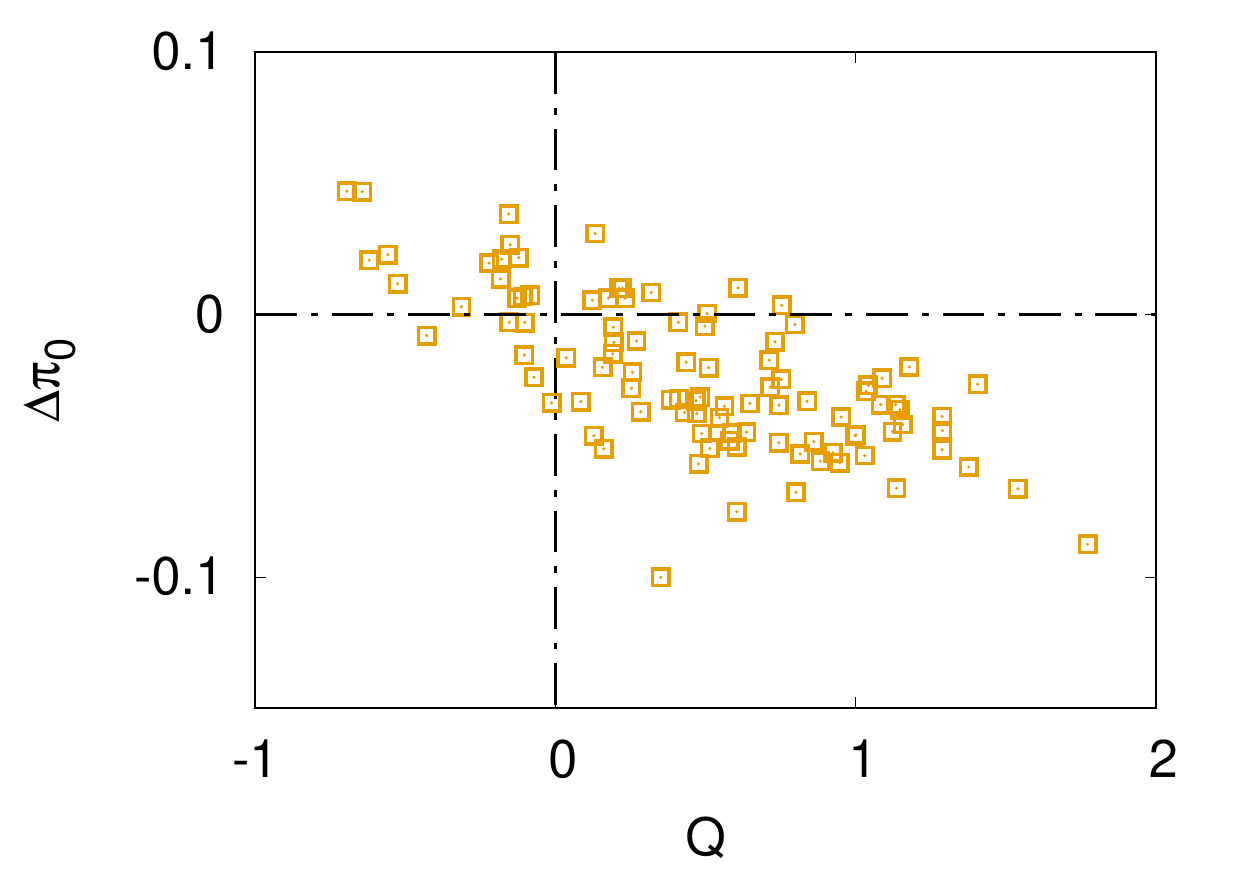}
\caption{Payoff difference between the resident population using Simple Standing
and its `slope mutant', which has the same $\beta$ and $\alpha(1,1,1)=1$ but
different slopes $\track{A_x}$ and $\track{A_y}$ and $\track{A_z}$.
Each point denotes a randomly generated mutant through the trilinear
interpolation among $\alpha(1,1,1)=1$ and seven random values $\alpha(0,0,0),
\alpha(0,0,1), \ldots, \alpha(1,1,0)$ within the unit interval.
The mutant norm occupies $10\%$ of the whole population whose size is $N=100$.
The horizontal axis shows the mutant's $Q$-value [Eq.~\eqref{eq:Q}], and the
vertical axis means its payoff difference $\Delta \pi_0$ with respect to the
resident norm after a sufficiently long time, e.g., $M/N \sim O(10^3)$.
As before, the game is
defined with $b=2$ and $c=1$, and the observation probability is $q=0.4$.
\track{Perception error and implementation error occur with probabilities
$e=0.1$ and $\gamma=0.1$, respectively.}
}
\label{fig:slope}
\end{center}
\end{figure}

Finally, we can numerically check the effect of a `slope mutant', which has
$\alpha(1,1,1)=1$ as a fixed point and the same behavioural rule as Simple
Standing but differs in the slopes $\track{A_x}$,
$\track{A_y}$ and $\track{A_z}$. To be more specific, let us assume that a mutant
norm occupies $10\%$ of the population whereas the rest of them are using Simple
Standing. The values of $\alpha(x,y,z)$ at the vertices of the three-dimensional
unit hypercube are randomly drawn from the unit interval, except for
$\alpha(1,1,1)=1$. Then, the trilinear interpolation is
used to construct the continuous assessment rule. According to our simulation
(Fig.~\ref{fig:slope}),
the performance of the mutant norm is strongly correlated with its
$Q$-value [Eq.~\eqref{eq:Q}]. Recall that the expression of $Q$ has been derived
in the context of recovery from small disagreement in a homogeneous population.
Figure~\ref{fig:slope} nevertheless suggests that it can also serve as a useful
indicator to tell if a minority of `slope mutants' will be competitive with
the resident norm, even when the difference between their assessment rules is
not necessarily small.

\section*{Summary and Discussion}

In summary, we have studied indirect reciprocity with private, noisy, and
incomplete information by extending the binary variables for
reputation and behaviour to continuous ones.
\track{%The extension to continuum is an idealization, abstracted from the fact
The extension to continuum is an idealization because it would impose an
excessive cognitive load to keep track
of others' reputations without discretization; nonetheless,
this abstraction allows us to overcome the fact
that the sharp dichotomy between good and bad is often found
insufficient in reporting an
assessment~\cite{alwin1997feeling,preston2000optimal,svensson2000comparison}.}
In particular, this formulation makes
it possible to check the role of sensitivity to new information
in judging others and adjusting our own behaviour.
That is, according to Eq.~\eqref{eq:bc}, the benefit-cost ratio of cooperation
should increase for stabilizing the cooperative initial state, if reputation is
insensitive to observed behaviour (low $\track{A_y}$) or if the level of
cooperation is insensitive to the recipient's reputation (low $\track{B_y}$).
At the same time, in contrast to the well-known condition
for indirect reciprocity akin to Hamilton's rule [Eq.~\eqref{eq:image}], we have
observed that incompleteness of information, controlled by $q<1$, mainly
affects the convergence toward a stationary state without altering the overall
conclusion.
This approach sheds light
on difference among the leading eight in their recovery
speeds from a single disagreement. Our analysis has identified the key
factor $\alpha_{1C0}$ in Table~\ref{table:eight}, i.e., how to assign
reputation to a well-reputed donor who chooses $C$ against an ill-reputed
recipient: If this choice is regarded as good according to $\alpha_{1C0}=1$,
making the assessment function $\alpha(x,y,z)$ insensitive to $z$, the recovery
can take place smoothly. As a result, we conclude that $\alpha$ should respond
to the donor's defection ($\track{A_y}>0$) but not necessarily to the players'
reputations (e.g., $\track{A_x} = \track{A_z} = 0$).
\track{A recent study also argues that
helping an ill-reputed player should be regarded as good to maintain stable
cooperation~\cite{okada2020two}.}
Such understanding of indirect reciprocity
in terms of sensitivity is important because, as is usual,
information processing through reputation has a trade-off between robustness and
sensitivity: One could underestimate new information and fail to adapt, or, one
could overestimate it and fail to distinguish noise from the signal. In
practice, the best way of assessment seems to be updating little by little upon
arrival of new information~\cite{tetlock2015superforecasting}, and such a
possibility is already incorporated in this continuum formulation.

It should be emphasized that our analysis has focused on local perturbation to
the existing norm. Therefore, our inequalities cannot be interpreted as a
condition for evolutionary stability against every possible mutant.
\track{
Moreover, although $\Delta \pi_0$ is found independent of $p$ in our
analysis, one should keep in mind that it results from a first-order theory so
that higher-order corrections generally show dependence on $p$.
If a mutant is sufficiently different from the resident,
then the first-order theory fails and the payoff difference may well
depend on $p$. For instance, if we think of a population consisting of L1 and L8
(Table~\ref{table:eight}),
we see that L1 is better off only when it
comprises the majority of the population (not shown).
Having said that, our local analysis can nevertheless provide a necessary
condition which will hold for stronger notions of stability as well.
}
We also believe that this locality assumption
is usually plausible in reality, considering that a
social norm is a complex construct that combines expectation and action in a
mutually reinforcing manner and thus resists change but small
ones~\cite{mackie2015social}. An empirical analysis shows that even orthographic
and lexical norms change so slowly that it takes centuries unless
intervened by a formal institution~\cite{amato2018dynamics}.
Another restriction in our analytic approach
is that the mutation is assumed to have
positive $\delta_1$ so that the mutant is not fully content with the initial
cooperative state. If two norms have $\delta_1=0$ in common and differ only by
slopes at the initial state, the first-order perturbation does not give a
definite answer as to their dynamics.
Having positive $\delta_1$ can be interpreted from a myopic player's point of
view as follows: A selfish player in a cooperating population may feel tempted
to devalue others' cooperation and reduce his or her own cost of cooperation
toward them. If our condition is met, however, such behaviour will eventually
be punished by the social norm.

``Maturity of mind is the capacity to endure uncertainty,'' says a maxim.
Although one lesson of the life is that we have to accept the grey area between
good and bad, reputation is still something that can be easily driven to
extremes, and what is worse is that it often goes in a different direction for
each
observer. Despite the theoretical achievement of indirect reciprocity, its real
difficulties are thus manifested in the problem of private assessment, noise,
and incomplete information. Our finding suggests that we can get a better grip
on indirect reciprocity by preparing reputational and behavioural scales with
finer gradations, which may be thought of as a form of systematic deliberation
to protect each other's reputation from rash judgement.

\section*{Methods}
\subsection*{Linear-order corrections}
\label{app:linear}

Equation~\eqref{eq:stationary} in the large-$N$ limit is written as follows:
\begin{eqnarray}
m_{00} &=& \frac{1}{N-1} \sum_{j\neq 0} \alpha_0 [m_{00}, \beta_0 (m_{00},
m_{0j}), m_{0j}] = \alpha_0 [m_{00}, \beta_0 (m_{00}, m_{01}),
m_{01}]\label{eq:m00}\\
m_{01} &=& \frac{1}{N-1} \sum_{j\neq 1} \alpha_0 [m_{01}, \beta_1 (m_{11},
m_{1j}), m_{0j}] \approx \alpha_0 [m_{01}, \beta_1 (m_{11}, m_{11}),
m_{01}]\label{eq:m01}\\
m_{10} &=& \frac{1}{N-1} \sum_{j\neq 0} \alpha_1 [m_{10}, \beta_0 (m_{00},
m_{0j}), m_{1j}] = \alpha_1 [m_{10}, \beta_0 (m_{00}, m_{01}),
m_{11}]\label{eq:m10}\\
m_{11} &=& \frac{1}{N-1} \sum_{j\neq 1} \alpha_1 [m_{11}, \beta_1 (m_{11},
m_{1j}), m_{1j}] \approx \alpha_1 [m_{11}, \beta_1 (m_{11}, m_{11}),
m_{11}].\label{eq:m11}
\end{eqnarray}
With $m_{00} = 1-\epsilon_{00}$, $m_{01} = 1-\epsilon_{01}$, and $m_{10} =
1-\epsilon_{10}$, Eq.~\eqref{eq:m01} becomes
\begin{eqnarray}
1-\epsilon_{01} &=& \alpha_0 (1-\epsilon_{01}, 1, 1-\epsilon_{01}) = \alpha
(1-\epsilon_{01}, 1, 1-\epsilon_{01}) - \delta (1-\epsilon_{01}, 1,
1-\epsilon_{01})\\
&\approx& \alpha(1,1,1) - \track{A_x} \epsilon_{01} - \track{A_z} \epsilon_{01} -
\delta(1,1,1) =
1 - \track{A_x} \epsilon_{01} - \track{A_z} \epsilon_{01} - \delta_1,
\end{eqnarray}
where $\alpha_\xi \equiv \left. \partial \alpha / \partial \xi
\right|_{(1,1,1)}$ and $\delta_1 \equiv \delta(1,1,1)$. Thus, we have
\begin{equation}
\epsilon_{01} \approx \left( 1-\track{A_x} - \track{A_z} \right)^{-1} \delta_1.
\label{eq:e01}
\end{equation}
Likewise,
\begin{eqnarray}
\beta_0 (1-\epsilon_{00}, 1-\epsilon_{01}) &=&
\beta (1-\epsilon_{00}, 1-\epsilon_{01}) -
\eta (1-\epsilon_{00}, 1-\epsilon_{01})\\
&\approx& 1 - \track{B_x} \epsilon_{00} - \track{B_y} \epsilon_{01} - \eta_1,
\end{eqnarray}
where $\beta_\lambda \equiv \left. \partial \beta / \partial \lambda
\right|_{(1,1)}$ and $\eta_1 \equiv \eta(1,1)$. Using this expression, we obtain
from Eq.~\eqref{eq:m10} the following:
\begin{eqnarray}
1-\epsilon_{10} &=& \alpha \left( 1-\epsilon_{10}, 1-\track{B_x} \epsilon_{00} -
\track{B_y} \epsilon_{01} -\eta_1, 1 \right) \\
&\approx& 1- \track{A_x} \epsilon_{10} - \track{A_y} \left( \track{B_x} \epsilon_{00} +
\track{B_y} \epsilon_{01} + \eta_1 \right),
\end{eqnarray}
which means
\begin{eqnarray}
\epsilon_{10} &=&
\frac{\track{A_y}}{1-\track{A_x}} (\track{B_x} \epsilon_{00} + \track{B_y}
\epsilon_{01} + \eta_1 )\\
&=& \frac{\track{A_y}}{1-\track{A_x}} \left[ \track{B_x} \epsilon_{00} + \track{B_y}
(1-\track{A_x} - \track{A_z})^{-1} \delta_1 + \eta_1 \right].
\label{eq:e10}
\end{eqnarray}
To get a closed-form expression for this, we need $\epsilon_{00}$ in
addition to $\epsilon_{01}$ [Eq.~\eqref{eq:e01}].
Thus, from Eq.~\eqref{eq:m00}, we derive
\begin{eqnarray}
1-\epsilon_{00}
&\approx& \alpha \left[ 1-\epsilon_{00}, \beta_0 (1-\epsilon_{00},
1-\epsilon_{01}), 1-\epsilon_{01} \right] - \delta_1\\
&\approx& \alpha \left( 1-\epsilon_{00}, 1-\track{B_x} \epsilon_{00}
- \track{B_y} \epsilon_{01} - \eta_1, 1-\epsilon_{01} \right) - \delta_1\\
&\approx& 1 - \track{A_x} \epsilon_{00} - \track{A_y} \left( \track{B_x} \epsilon_{00} +
\track{B_y} \epsilon_{01} + \eta_1 \right) - \track{A_z} \epsilon_{01} - \delta_1,
\end{eqnarray}
which gives
\begin{eqnarray}
\epsilon_{00} &=&
\frac{1}{1-\track{A_x} - \track{A_y} \track{B_x}} \left[ (\track{A_y}
\track{B_y} + \track{A_z}) \epsilon_{01} + \track{A_y} \eta_1 + \delta_1 \right]\\
&=& \frac{1}{1-\track{A_x} - \track{A_y} \track{B_x}} \left[ \frac{\track{A_y}
\track{B_y} + \track{A_z}}{1-\track{A_x} - \track{A_z}} \delta_1  + \track{A_y} \eta_1 +
\delta_1 \right],
\label{eq:e00}
\end{eqnarray}
where we have used Eq.~\eqref{eq:e01}. By substituting Eq.~\eqref{eq:e00} into
Eq.~\eqref{eq:e10}, we can write $\epsilon_{10}$ explicitly.

\subsection*{Finite fraction of mutants}
\label{app:fraction}

If a mutant norm occupies a finite fraction $p$, Eqs.~\eqref{eq:m00} to
\eqref{eq:m11} are generalized to
\begin{eqnarray}
m_{00} &=& p \alpha_0 [m_{00}, \beta_0(m_{00}, m_{00}), m_{00}] + \bar{p}
\alpha_0 [m_{00}, \beta_0(m_{00}, m_{01}), m_{01}]\\
m_{01} &=& p \alpha_0 [m_{01}, \beta_1(m_{11}, m_{10}), m_{00}] + \bar{p}
\alpha_0 [m_{01}, \beta_1(m_{11}, m_{11}), m_{01}]\\
m_{10} &=& p \alpha_1 [m_{10}, \beta_0(m_{00}, m_{00}), m_{10}] + \bar{p}
\alpha_1 [m_{10}, \beta_0(m_{00}, m_{01}), m_{11}]\\
m_{11} &=& p \alpha_1 [m_{11}, \beta_1(m_{11}, m_{10}), m_{10}] + \bar{p}
\alpha_1 [m_{11}, \beta_1(m_{11}, m_{11}), m_{11}],
\end{eqnarray}
where $\bar{p} \equiv 1-p$.
Through linearisation, the above equations are rewritten as
\begin{eqnarray}
1-\epsilon_{00} &\approx& p [1-\track{A_x} \epsilon_{00} - \track{A_y} (\track{B_x}
\epsilon_{00} + \track{B_y} \epsilon_{00} + \eta_1) - \track{A_z} \epsilon_{00} -
\delta_1]\nonumber\\
&+& \bar{p} [1-\track{A_x} \epsilon_{00} - \track{A_y} (\track{B_x} \epsilon_{00} +
\track{B_y} \epsilon_{01} + \eta_1) - \track{A_z} \epsilon_{01} - \delta_1]\\
1-\epsilon_{01} &\approx& p [1-\track{A_x} \epsilon_{01} - \track{A_y} (\track{B_x}
\epsilon_{11} + \track{B_y} \epsilon_{10}) - \track{A_z} \epsilon_{00} -
\delta_1]\nonumber\\
&+& \bar{p} [1-\track{A_x} \epsilon_{01} - \track{A_y} (\track{B_x} \epsilon_{11} +
\track{B_y} \epsilon_{11}) - \track{A_z} \epsilon_{01} - \delta_1]\\
1-\epsilon_{10} &\approx& p [1-\track{A_x} \epsilon_{10} - \track{A_y} (\track{B_x}
\epsilon_{00} + \track{B_y} \epsilon_{00} + \eta_1) - \track{A_z}
\epsilon_{10}]\nonumber\\
&+& \bar{p} [1-\track{A_x} \epsilon_{10} - \track{A_y} (\track{B_x} \epsilon_{00} +
\track{B_y} \epsilon_{01} + \eta_1) - \track{A_z} \epsilon_{11}]\\
1-\epsilon_{11} &\approx& p [1-\track{A_x} \epsilon_{11} - \track{A_y} (\track{B_x}
\epsilon_{11} + \track{B_y} \epsilon_{10}) - \track{A_z}
\epsilon_{10}]\nonumber\\
&+& \bar{p} [1-\track{A_x} \epsilon_{11} - \track{A_y} (\track{B_x} \epsilon_{11} +
\track{B_y} \epsilon_{11}) - \track{A_z} \epsilon_{11}].
\end{eqnarray}
After some algebra, we find
\begin{eqnarray}
\epsilon_{00} &=&
\frac{\delta_1 \left\{ \track{A_x}^2+\track{A_x} (\track{A_y}
\track{B_x}+\track{A_z}-2)-\bar{p}\track{A_y}^2 \track{B_x}
\track{B_y}-\bar{p}\track{A_y}^2 \track{B_y}^2+\track{A_z} [\track{A_y}
(p\track{B_x}-\bar{p}\track{B_y})-1]-\track{A_y} \track{B_x}+1\right\}}
{(1-\track{A_x}-\track{A_z})
(1-\track{A_x}-\track{A_y} \track{B_x}) (1-\track{A_x}-\track{A_y} \track{B_x}-\track{A_y}
\track{B_y}-\track{A_z})}\nonumber\\
&&+\frac{\track{A_y} \eta_1
(1-\track{A_x}-\track{A_z}) (1-\track{A_x}-\track{A_y} \track{B_x}
-\bar{p}\track{A_y} \track{B_y}-\bar{p}\track{A_z})}
{(1-\track{A_x}-\track{A_z})
(1-\track{A_x}-\track{A_y} \track{B_x}) (1-\track{A_x}-\track{A_y} \track{B_x}-\track{A_y}
\track{B_y}-\track{A_z})}\label{eq:ep00}\\
\epsilon_{01} &=&
\frac{\track{A_y} \eta_1 p (1-\track{A_x}-\track{A_z}) (\track{A_y}
\track{B_y}+\track{A_z})}
{(1-\track{A_x}-\track{A_z})
(1-\track{A_x}-\track{A_y} \track{B_x}) (1-\track{A_x}-\track{A_y} \track{B_x}-\track{A_y}
\track{B_y}-\track{A_z})}\nonumber\\
&&
+\frac{\delta_1 \left[\track{A_x}^2+\track{A_x} (2 \track{A_y}
\track{B_x}+\track{A_y} \track{B_y}+\track{A_z}-2)+\track{A_y}^2 \track{B_x}^2+\track{A_y}^2 \track{B_x}
\track{B_y} p+\track{A_y}^2 \track{B_x} \track{B_y}+\track{A_y}^2 \track{B_y}^2 p
\right]}
{(1-\track{A_x}-\track{A_z})
(1-\track{A_x}-\track{A_y} \track{B_x}) (1-\track{A_x}-\track{A_y} \track{B_x}-\track{A_y}
\track{B_y}-\track{A_z})}\nonumber\\
&&
+\frac{\delta_1 \left\{
\track{A_z} [\track{A_y}
(p\track{B_x}+\track{B_x}+p\track{B_y})-1]-2 \track{A_y} \track{B_x}-\track{A_y}
\track{B_y}+1\right\}}
{(1-\track{A_x}-\track{A_z})
(1-\track{A_x}-\track{A_y} \track{B_x}) (1-\track{A_x}-\track{A_y} \track{B_x}-\track{A_y}
\track{B_y}-\track{A_z})}\label{eq:ep01}\\
\epsilon_{10} &=&
\frac{\track{A_y} (1-\track{A_x}-\track{A_y} \track{B_x}-\bar{p}\track{A_y}
\track{B_y}-\bar{p}\track{A_z}) [\eta_1 (1-\track{A_x}-\track{A_z})+(\track{B_x}
+\track{B_y}) \delta_1]}
{(1-\track{A_x}-\track{A_z})
(1-\track{A_x}-\track{A_y} \track{B_x}) (1-\track{A_x}-\track{A_y} \track{B_x}-\track{A_y}
\track{B_y}-\track{A_z})}\label{eq:10}\\
\epsilon_{11} &=&
\frac{\track{A_y} p (\track{A_y} \track{B_y}+\track{A_z}) (\eta_1
(1-\track{A_x}-\track{A_z})+(\track{B_x} +\track{B_y}) \delta_1)}
{(1-\track{A_x}-\track{A_z})
(1-\track{A_x}-\track{A_y} \track{B_x}) (1-\track{A_x}-\track{A_y} \track{B_x}-\track{A_y}
\track{B_y}-\track{A_z})},\label{eq:ep11}
\end{eqnarray}
from which one can reproduce the previous results [Eqs.~\eqref{eq:epsilon00} to
\eqref{eq:epsilon10}] by taking the limit of $p \to 0$. The
denominators seem to require another inequality in addition to
Eqs.~\eqref{eq:ineq1} and \eqref{eq:ineq2}, that is,
\begin{equation}
\track{A_x} + \track{A_z} + \track{A_y} (\track{B_x} + \track{B_y}) < 1,
\label{eq:spurious}
\end{equation}
which is equivalent to Eq.~\eqref{eq:Q}.
Recall that the continuous versions of the leading eight always have $\track{A_y} =
\track{B_y} = 1$ and $\track{A_x} = \track{B_x} = 0$ in common, which means that they all
violate this inequality. However, in practice, no singularity arises for Simple
Standing if higher-order corrections are included, and even the
second-order calculation agrees moderately well with numerical results.

The payoff earned by a mutant is calculated as
\begin{eqnarray}
\pi_0 &=& b [p \beta_0 (m_{00}, m_{00}) + (1-p) \beta_1(m_{11}, m_{10})]
\nonumber\\
&&-c [p \beta_0 (m_{00}, m_{00}) + (1-p) \beta_0(m_{00}, m_{01})]\\
&\approx& b [p (1-\track{B_x} \epsilon_{00} - \track{B_y} \epsilon_{00} - \eta_1)
+ (1-p)(1-\track{B_x} \epsilon_{11} - \track{B_y} \epsilon_{10})]\nonumber\\
&& -c [p (1-\track{B_x} \epsilon_{00} - \track{B_y} \epsilon_{00} - \eta_1) + (1-p)
[1-\track{B_x} \epsilon_{00} - \track{B_y} \epsilon_{01} - \eta_1]],
\end{eqnarray}
whereas a resident player earns
\begin{eqnarray}
\pi_1 &=& b [p \beta_0 (m_{00}, m_{01}) + (1-p) \beta_1(m_{11}, m_{11})]
\nonumber\\
&&-c [p \beta_1 (m_{11}, m_{10}) + (1-p) \beta_1(m_{11}, m_{11})]\\
&\approx& b [p (1-\track{B_x} \epsilon_{00} - \track{B_y} \epsilon_{01} - \eta_1)
+ (1-p)(1-\track{B_x} \epsilon_{11} - \track{B_y} \epsilon_{11})]\nonumber\\
&& -c [p(1-\track{B_x} \epsilon_{11} - \track{B_y} \epsilon_{10}) + (1-p) (1-\track{B_x}
\epsilon_{11} - \track{B_y} \epsilon_{11})].
\end{eqnarray}
If we plug Eqs.~\eqref{eq:ep00} to \eqref{eq:ep11} here,
the payoff difference $\Delta \pi_0 = \pi_0 - \pi_1$ becomes identical to
Eq.~\eqref{eq:dpi0} with no dependence on $p$.

\subsection*{Second-order corrections}
\label{app:second}

We assume that $\delta$, $\eta$, as well as their partial derivatives, and
$\epsilon_{ij}$'s are small parameters of the same order of magnitude. The
second-order perturbation for $\beta_1$ can thus be written as follows:
\begin{eqnarray}
\beta_1 (m_{11}, m_{1j}) &=& \beta(1-\epsilon_{11}, 1-\epsilon_{1j})\\
&\approx& 1 - \track{B_x} \epsilon_{11} - \track{B_y} \epsilon_{1j} +
\frac{1}{2} \track{B_{xx}} \epsilon_{11}^2 + \track{B_{xy}} \epsilon_{11} \epsilon_{1j}
+ \frac{1}{2} \track{B_{yy}} \epsilon_{1j}^2 \\
&\equiv& 1 - \kappa_1.
\end{eqnarray}
Here, we write $\kappa_1 \equiv \kappa_1^{(1)} + \kappa_1^{(2)}$, where
$\kappa_1^{(1)} \equiv \track{B_x} \epsilon_{11} + \track{B_y}
\epsilon_{1j}$ and $\kappa_1^{(2)} \equiv -
\left( \frac{1}{2} \track{B_{xx}} \epsilon_{11}^2 + \track{B_{xy}} \epsilon_{11}
\epsilon_{1j} + \frac{1}{2} \track{B_{yy}} \epsilon_{1j}^2 \right)$ are first- and
second-order corrections, respectively, \track{and $B_{\mu\nu} \equiv \left.
\partial^2 \beta /\partial \mu \partial \nu \right|_{(1,1)}$}.
Likewise,
\begin{eqnarray}
\beta_0 (m_{00}, m_{0j}) &=& \beta(m_{00}, m_{0j}) - \eta(m_{00}, m_{0j})\\
&=& \beta(1-\epsilon_{00}, 1-\epsilon_{0j}) - \eta(1-\epsilon_{00},
1-\epsilon_{0j})\\
&\approx& \left(1 - \track{B_x} \epsilon_{00} - \track{B_y} \epsilon_{0j} +
\frac{1}{2} \track{B_{xx}} \epsilon_{00}^2 + \track{B_{xy}} \epsilon_{00}
\epsilon_{0j} + \frac{1}{2} \track{B_{yy}} \epsilon_{0j}^2 \right)\nonumber\\
&&- \left(\eta_1 - \eta_x \epsilon_{00} - \eta_y \epsilon_{0j} \right)\\
&\equiv& 1 - \kappa_0,
\end{eqnarray}
where $\kappa_0 \equiv \kappa_0^{(1)} + \kappa_0^{(2)}$ with $\kappa_0^{(1)}
\equiv \track{B_x} \epsilon_{00} + \track{B_y} \epsilon_{0j} + \eta_1$ and
$\kappa_0^{(2)} \equiv -\left( \frac{1}{2} \track{B_{xx}} \epsilon_{00}^2
+ \track{B_{xy}} \epsilon_{00} \epsilon_{0j} + \frac{1}{2} \track{B_{yy}}
\epsilon_{0j}^2 \right) - (\eta_x \epsilon_{00} + \eta_y \epsilon_{0j})$.

The second-order perturbation for $\alpha_1$ is also straightforward:
\begin{eqnarray}
\alpha_1 [m_{1i}, \beta_i (m_{ii}, m_{ij}), m_{1j}]
&\approx& \alpha(1-\epsilon_{1i}, 1-\kappa_i, 1-\epsilon_{1j})\\
&\approx& 1 - \track{A_x} \epsilon_{1i} - \track{A_y} \kappa_i - \track{A_z}
\epsilon_{1j}
+ \frac{1}{2} \track{A_{xx}} \epsilon_{1i}^2 + \frac{1}{2} \track{A_{yy}}
\left(\kappa_i^{(1)} \right)^2 + \frac{1}{2} \track{A_{zz}}
\epsilon_{1j}^2\nonumber\\
&&+ \track{A_{xy}} \epsilon_{1i} \kappa_i^{(1)}
+ \track{A_{yz}} \kappa_i^{(1)} \epsilon_{1j}
+ \track{A_{zx}} \epsilon_{1i} \epsilon_{1j},
\end{eqnarray}
\track{
where $A_{\mu \nu} \equiv \left. \partial^2 \alpha / \partial\mu \partial\nu
\right|_{(1,1,1)}$,
}
and similarly,
\begin{eqnarray}
\alpha_0 [m_{0i}, \beta_i (m_{ii}, m_{ij}), m_{0j}]
&\approx& \alpha(1-\epsilon_{0i}, 1-\kappa_i, 1-\epsilon_{0j}) -
\delta(1-\epsilon_{0i}, 1-\kappa_i, 1-\epsilon_{0j})\\
&\approx& \left[ 1 - \track{A_x} \epsilon_{0i} - \track{A_y} \kappa_i - \track{A_z}
\epsilon_{0j}
+ \frac{1}{2} \track{A_{xx}} \epsilon_{0i}^2 + \frac{1}{2} \track{A_{yy}}
\left(\kappa_i^{(1)} \right)^2 + \frac{1}{2} \track{A_{zz}}
\epsilon_{0j}^2 \right.\nonumber\\
&& \left. + \track{A_{xy}} \epsilon_{0i} \kappa_i^{(1)}
+ \track{A_{yz}} \kappa_i^{(1)} \epsilon_{0j}
+ \track{A_{zx}} \epsilon_{0i} \epsilon_{0j}
\nonumber \right]\\
&&- \left( \delta_1 - \delta_x \epsilon_{0i} - \delta_y \kappa_i^{(1)} -
\delta_z \epsilon_{0j} \right).
\end{eqnarray}

%\bibliography{rep}

\section*{Acknowledgements}
Y.M. acknowledges support from Japan Society for the Promotion of Science (JSPS)
(JSPS KAKENHI; Grant no. 18H03621 and Grant no. 21K03362).
S.K.B. acknowledges support by Basic Science Research Program through the
National Research Foundation of Korea (NRF) funded by the Ministry of Education
(NRF-2020R1I1A2071670).
We appreciate the  APCTP for its hospitality during completion of this work.

\section*{Author contributions statement}
S.L. carried out computation and analysed the results.
Y.M. verified the method and reviewed the manuscript.
S.K.B. conceived the work and wrote the manuscript.

\section*{Additional information}

The authors declare no competing financial interests.

\section*{Data Availability}
\track{The source code for this study is available at
\url{https://github.com/yohm/sim_game_continuous_reputation}.}

\end{document}